\documentclass[prd, twocolumn, superscriptaddress,floatfix, nofootinbib, preprintnumbers]{revtex4-2}
\usepackage[utf8]{inputenc}
\usepackage[colorlinks=true,citecolor=blue]{hyperref}
\usepackage{amssymb}
\usepackage{amsmath}
\usepackage{graphicx}
\usepackage{footmisc}
\usepackage{url}
\usepackage{multirow}
\usepackage{makecell}
\usepackage{enumerate}
\usepackage{hhline}
\usepackage{comment}
\usepackage{nicefrac}
\usepackage{placeins}

\def\beq{\begin{equation}}
\def\eeq{\end{equation}}
\newcommand{\bea}{\begin{eqnarray}\begin{aligned}}
\newcommand{\eea}{\end{aligned}\end{eqnarray}}
\newcommand{\cathode}{C\textsc{athode}}
\newcommand{\lacathode}{LaC\textsc{athode}}
\newcommand{\cwola}{CW\textsc{o}L\textsc{a}}
\newcommand{\anode}{A\textsc{node}}
\newcommand{\curtains}{C\textsc{urtain}s}

\begin{document}

\title{Resonant anomaly detection without background sculpting}

\author{Anna Hallin}
\email{anna.hallin@rutgers.edu}
\affiliation{NHETC, Dept.\ of Physics and Astronomy, Rutgers University, Piscataway, NJ 08854, USA}

\author{Gregor Kasieczka}
\email{gregor.kasieczka@uni-hamburg.de}
\affiliation{Institut f\"{u}r Experimentalphysik, Universit\"{a}t Hamburg, 22761 Hamburg, Germany}
\affiliation{Center for Data and Computing in Natural Sciences (CDCS), 22607 Hamburg, Germany}

\author{Tobias Quadfasel}
\email{tobias.quadfasel@uni-hamburg.de}
\affiliation{Institut f\"{u}r Experimentalphysik, Universit\"{a}t Hamburg, 22761 Hamburg, Germany}

\author{David Shih}
\email{shih@physics.rutgers.edu}
\affiliation{NHETC, Dept.\ of Physics and Astronomy, Rutgers University, Piscataway, NJ 08854, USA}

\author{Manuel Sommerhalder}
\email{manuel.sommerhalder@uni-hamburg.de}
\affiliation{Institut f\"{u}r Experimentalphysik, Universit\"{a}t Hamburg, 22761 Hamburg, Germany}

\begin{abstract}
We introduce a new technique named \textit{Latent CATHODE} (\lacathode{}) for performing ``enhanced bump hunts”, a type of resonant anomaly search that combines conventional one-dimensional bump hunts with a model-agnostic anomaly score in an auxiliary feature space where potential signals could also be localized.
The main advantage of \lacathode{} over existing methods is that it provides an anomaly score that is well behaved when evaluating it beyond the signal region, which is essential to prevent the sculpting of background distributions in the bump hunt. 
\lacathode{} accomplishes this by constructing the anomaly score directly in the latent space learned by a conditional normalizing flow trained on sideband regions. 
We demonstrate the superior stability and comparable performance of \lacathode{} for enhanced bump hunting in an illustrative toy example as well as on the LHC Olympics R\&D dataset.
\end{abstract}

\maketitle

\section{Introduction}

Despite countless searches for new physics at the LHC, so far no evidence for physics beyond the Standard Model was found. The vast majority of these searches are model specific, motivated by and optimized for particular scenarios and particle spectra. Recently there has been much interest in the possibility that new physics could be present in the data but we simply have not searched in the right places yet. This has led to an enormous activity in developing new methods for model-agnostic searches at the LHC (see e.g.~\cite{Kasieczka:2021xcg,Aarrestad:2021oeb} for recent community overviews of anomaly detection and \cite{Karagiorgi:2021ngt} for a more general overview of machine learning methods to search for new physics).

One promising class of approaches can be referred to as ``enhanced bump hunts", where the idea is to upgrade a standard one-dimensional bump hunt, e.g.\ in an invariant mass $m$\footnote{We will use $m$ for illustration in this text, but all features in which the signal is resonant and the background is smooth can be used~\cite{Kasieczka:2021tew}.},
to a multivariate setting. This is achieved by including an anomaly score $R(x)$ learned from auxiliary features $x\in {\Bbb R}^d$ where the signal may also be localized, but in an a priori unknown way.

In general, enhanced bump hunts follow these steps:
\begin{enumerate}[i)]
\item Designate nonoverlapping signal region (SR) and sidebands (SB) in $m$.
\item Derive an anomaly score $R(x)$ and select events that pass a threshold value $R(x) > R_c$.
\item Fit a suitable (e.g.\ falling spectrum) background-only function to the selected events in the SB.
\item Compare the background-only prediction from step iii)
to data in the SR and derive limits or claim discovery.
\end{enumerate}

Methods for enhanced bump hunts include those constructed using 
autoencoders~\cite{Heimel:2018mkt,Farina:2018fyg} or based on weak supervision~\cite{Collins:2018epr,Collins:2019jip}.
While weak supervision allows the construction (in an ideal case) of a provably optimal anomaly score---see Section~\ref{sec:bumphunts} for details---correlations between the bump hunt feature and auxiliary features can spoil these methods.
This observation has motivated the development of a number of new techniques that aim to improve the sensitivity and stability of anomaly detection in the presence of correlations~\cite{Nachman:2020lpy,Andreassen:2020nkr,1815227,Hallin:2021wme,Raine:2022hht}. 
In particular, the recently proposed \cathode{}~\cite{Hallin:2021wme} and \curtains{}~\cite{Raine:2022hht} techniques have been demonstrated to achieve close-to-optimal signal sensitivity, even in the presence of correlations between features.
 
This paper is concerned with another issue that has received less attention but still might spoil the practical application of enhanced bump hunts: background sculpting. The enhanced bump hunting procedure outlined above can only work if the cut introduced in ii) does not sculpt the background (i.e. introduce artificial bumps in the background-only $m$ spectrum).
Alas, state-of-the-art protocols like \cathode{} and \curtains{} have no built-in measures to prevent such sculpting.
Even worse, the anomaly score of these approaches is only derived for the SR, leading to potentially unpredictable extrapolation behavior elsewhere.

The scope of this paper is to clearly identify this sculpting issue and to provide a viable solution.
In Sec.~\ref{sec:method}, we first discuss enhanced bump hunt strategies and then introduce the novel \lacathode{} approach.
Section~\ref{sec:toy}  uses an analytic toy model to illustrate the problem and shows that  correlations
between $m$ and the auxiliary features are the root cause of background sculpting.
It also shows that \lacathode{} indeed successfully mitigates this issue.
Section~\ref{sec:LHCO} reiterates these points, but in the context of the more physically motivated LHC Olympics R\&D dataset \cite{gregor_kasieczka_2019_6466204}. Section~\ref{sec:conclusions} concludes this work.

\section{Method}
\label{sec:method}

\subsection{Existing strategies for enhanced bump hunts}

\label{sec:bumphunts}

According to the Neyman-Pearson lemma~\cite{neyman1933ix}, the provably optimal anomaly score for any model-agnostic search would be:
\beq\label{eq:Rideal}
R(x)={p_{data}(x)\over p_{bg}(x)}
\eeq
where $p_{data}(x)$ and $p_{bg}(x)$ are the probability densities of the data and the background respectively.
Of course, in practice we never have access to this likelihood ratio, since the probability densities of data and background are in general intractable. At best, one could hope for a large number of {\it samples} drawn from the data and true background distributions; then one could approximate $R(x)$ with a classifier trained on these samples. We will refer to this approximation of (\ref{eq:Rideal}) as the ``idealized anomaly detector" throughout.

Since it is generally not possible to draw samples from the true $p_{bg}(x)$ in a realistic anomaly search scenario, we can at best approximate this idealized case either with simulations or in a data-driven way. The focus here will be on the latter strategy.  

The challenge then is to
obtain a high-quality estimate for $p_{bg}(x)$ from data, e.g. by interpolating from sidebands (SB) in $m$ into a signal region (SR), and use weak supervision to obtain an anomaly score $R(x)$.
{\it As long as a cut on $R(x)>R_c$ does not sculpt the $m$ distribution}, one can combine this cut with the 1D bump hunt in $m$ to greatly enhance the significance of the signal over the background. 
 
In the original enhanced bump hunt method, called \cwola{}-Hunting~\cite{Collins:2019jip}, $R(x)$ comes from a SR vs SB classifier. This works as long as the features $x$ and $m$ are statistically independent in the background (i.e.\ the $x$ features are distributed identically in the SR and the SB for the background). This also ensures that $R(x)>R_c$ will not sculpt the $m$ distribution. Using these properties, the full enhanced bump hunt search strategy using \cwola{}-Hunting was successfully demonstrated on toy simulation data ~\cite{Collins:2019jip, Collins:2021nxn}, and then implemented on actual data by the ATLAS Collaboration in~\cite{collaboration2020dijet}. 

However, it can be challenging to ensure that $x$ and $m$ are independent in the background. Even a small correlation can degrade or destroy the sensitivity of \cwola{} Hunting to anomalies. This has motivated the development of alternative approaches that are more robust to correlations. 
\begin{itemize}
    
    \item In \anode{}~\cite{Nachman:2020lpy}, one learns $p_{data}(x)$ and $p_{bg}(x)$ using conditional density estimators trained on the data with $m\in SR$ and with $m\in SB$; the latter are automatically interpolated in $m$ into the SR, which alleviates the problem with correlations between $x$ and $m$. It was shown in~\cite{Hallin:2021wme} that in the presence of correlations between $x$ and $m$, the signal sensitivity of \anode{} is robust while that of \cwola{}-Hunting collapses. 
    
    \item In \cathode{}~\cite{Hallin:2021wme}, one learns $p_{bg}(x)$ using the SB density estimator just as in \anode{}. However, instead of the second SR density estimator (which will be more difficult to learn as it must also capture the tiny deviations from the smooth $p_{bg}(x)$ from a small localized signal), one samples from $p_{bg}(x)$ in the SR, and trains a classifier (as in \cwola{}-Hunting) between the data and the synthetic background samples. \cathode{} thereby captures the best of both \anode{} and \cwola{}-Hunting, achieving a signal sensitivity that is nearly optimal and yet robust to correlations between $x$ and $m$. 
    
    \item Finally, the \curtains{}~\cite{Raine:2022hht} protocol operates similar to \cathode{}, with the main difference that conditional invertible neural networks (cINNs) are used to map background examples from the SB into the SR.

\end{itemize}

\subsection{The problem of background sculpting}

\begin{figure}[htb!]
    \centering
    \includegraphics[width=0.45\textwidth]{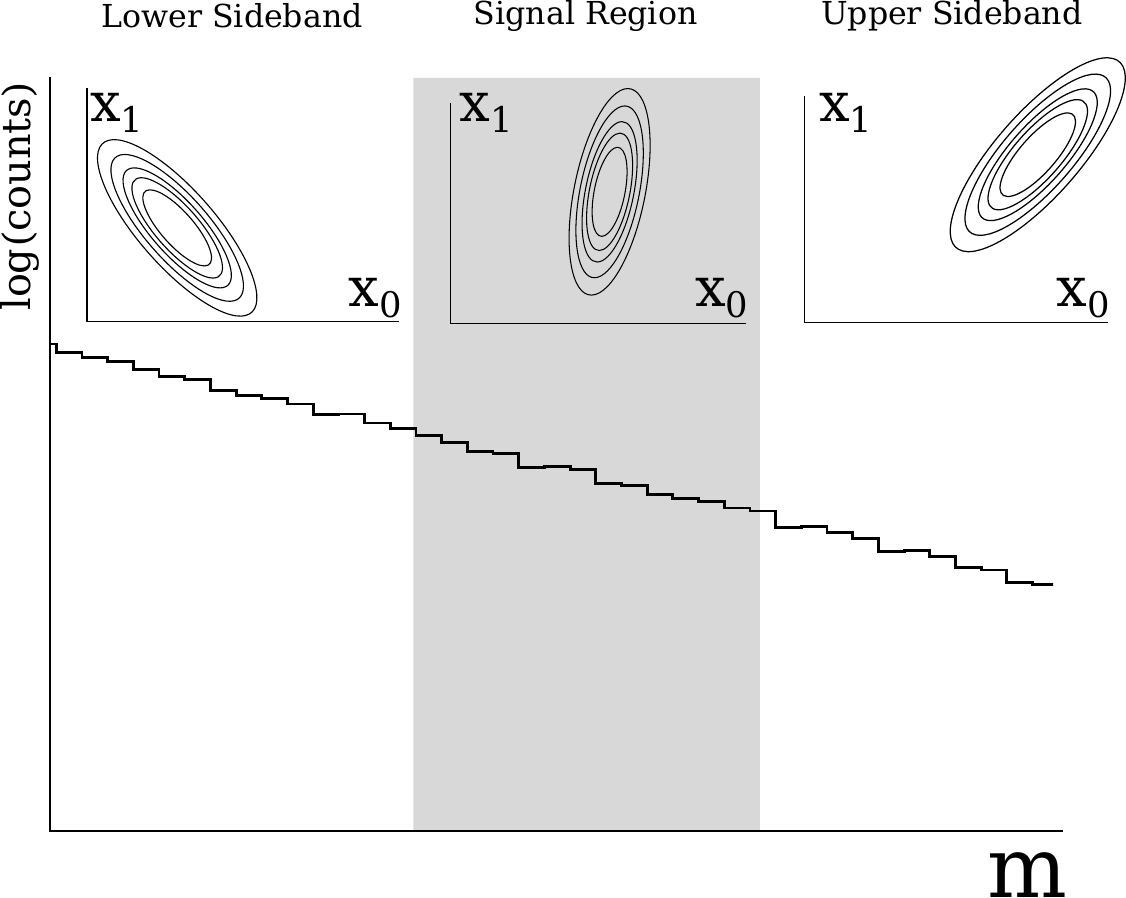}
    \caption{Illustration of the correlation of (hypothetical) input features $x_{0}$ and $x_{1}$ with $m$ in the background. This figure describes the situation mentioned in the text, where one can clearly observe that the background distributions of input features $x_{0}$ and $x_{1}$ change dramatically from the lower sideband (low $m$) to the upper sideband (high $m$) and thus are strongly correlated with $m$.}
    \label{fig:correlations}
\end{figure}

So far, apart from \cwola{}-Hunting, the majority of the effort has been invested in exploring data-driven approaches to learn $R(x)$ as accurately as possible from sidebands, while much less attention has been paid to the issue of background sculpting.
However, signal sensitivity is not the only component of a successful new physics search; background estimation is also essential. In the presence of correlations between $x$ and $m$ in the background events, one must also show that $R(x)$, even if ideal, does not sculpt the background $m$ distribution around the signal region, which would prevent background estimation via the 1D bump hunt. See Fig.~\ref{fig:correlations} for an illustration of such correlated input features.

Note that, in any complete enhanced bump hunt strategy, \textit{two} data-driven background estimations must take place: 
\begin{enumerate}
    \item An interpolation of the learned $p_{bg}(x)$ from SB to SR in order to construct $R(x)$.
    \item After cutting on $R(x)>R_c$, we proceed with the usual 1D bump hunt: an interpolation in the $m$ distribution from SB to SR (e.g.\ by fitting a suitable functional form to the data excluding the SR). 
\end{enumerate}

This work is concerned with ensuring the robustness of the second estimation.
We will demonstrate---using both a simple analytic toy model and examples drawn from the LHC Olympics 2020 R\&D dataset~\cite{gregor_kasieczka_2019_6466204}---that in the presence of correlations between $x$ and $m$, cutting on the learned $R(x)$ can result in significant sculpting of the $m$ distribution. This can be understood by the fact that $R(x)$ must be a more-or-less smooth function of $x$, so any correlations of $m$ with $x$ will be inherited by $R(x)$. Furthermore, $R(x)$ was learned using events in the SR, so it has to be {\it extrapolated} from SR to SB in order to apply the threshold $R(x)>R_{c}$ everywhere. This extrapolation could lead to unpredictable effects, including sculpting, especially in the presence of strong correlations between $x$ and $m$.

\begin{figure*}[htb!]
    \centering
    \includegraphics[width=\textwidth]{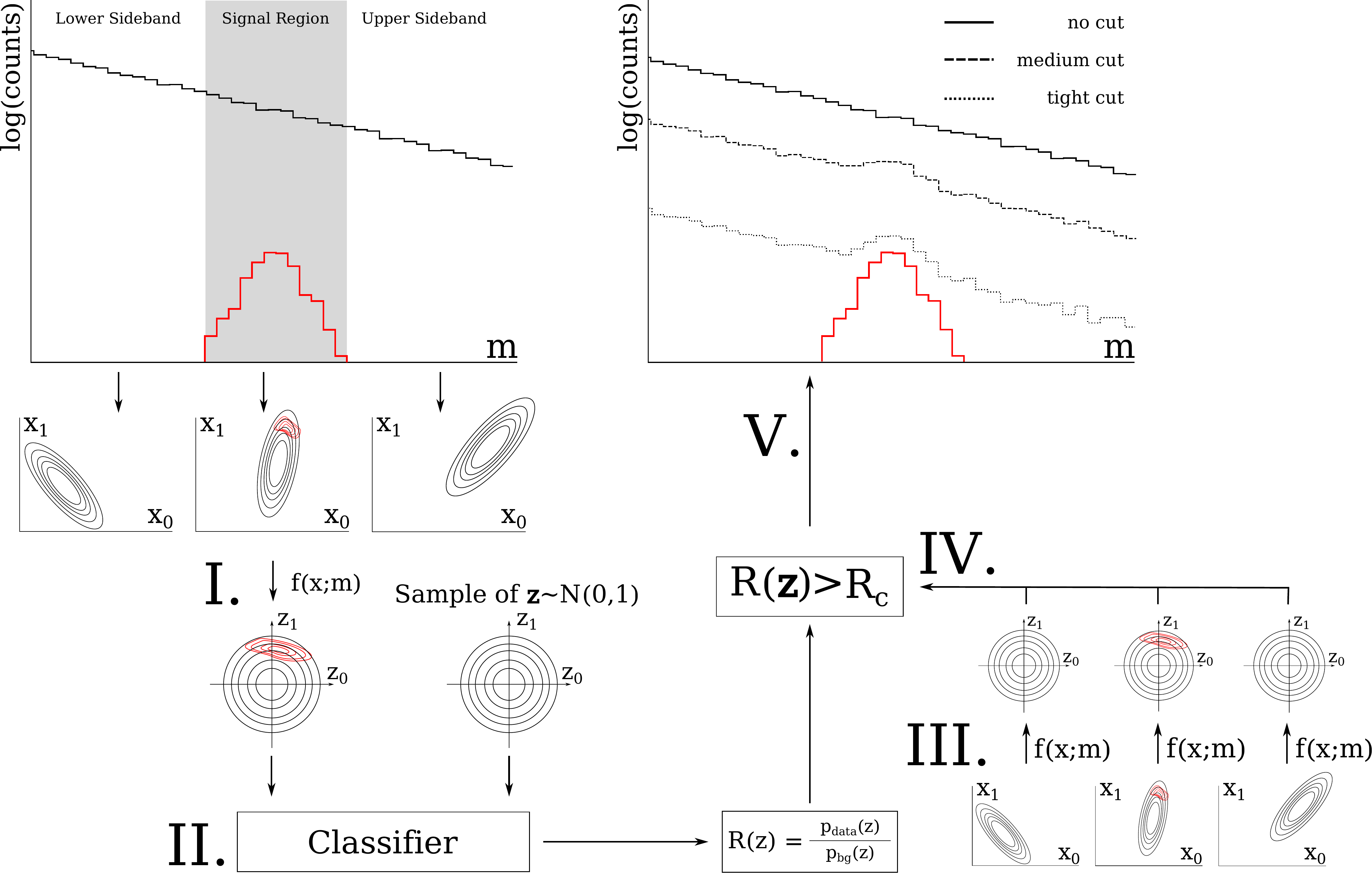}
    \caption{Flowchart describing the different steps of the proposed \lacathode{} method. First, one maps the data in the SR (denoted by the vertical gray band) to the latent space (I), and trains a classifier $R(z)$ to distinguish data from the background which follows the normal distribution (II). Then one maps all the data (in both SR and SB) to the latent space (III) and passes this through $R(z)$, selecting only those events above some threshold $R(z)>R_c$ (IV). Finally, one plots the $m$ distribution of the surviving events, and looks for a bump in the SR that would signify the presence of new physics (V). }
    \label{fig:flowchart}
\end{figure*}

\subsection{LaCATHODE to the rescue}

After identifying the issue that leads to sculpting of the background $m$ distribution, we present a solution to the problem, which is outlined in Fig.~\ref{fig:flowchart} and described in the following. We call our new approach {\it Latent CATHODE} or \lacathode{} for short, because it is closely derived from the \cathode{} method. 

The solution actually lies at the heart of the \cathode{} method: the SB density estimator is a conditional normalizing flow, which is an invertible map $f$ from data space $x$ to a latent space $z$, for every value of $m$:
\beq
z=f(x;m)
\eeq
The background events in the latent space $z$ are supposed to follow a simple prespecified base distribution, which we take to be the unit normal distribution ${\mathcal N}(\mu=0,\sigma=1)^d$ for concreteness. 

The idea of \lacathode{} is to train the classifier between SR data and SR background in the latent space $z$ instead of in the physical feature space $x$. 
Since $f$ maps $x$ to the same latent space for every value of $m$, working in the latent space has the effect of decorrelating the data from $m$ in the background, which should eliminate the problem of sculpting. In other words, since the $z$ space is always the same for every $m$, no extrapolation is needed to evaluate $R(z)$ outside the SR where it was learned.\footnote{A closely related approach would be to use the invertible map $f$ {\it twice} to decorrelate $x$ from $m$ in the SB regions: $x\to x' = f^{-1}(f(x,m),m_0)$ for some suitable choice of $m_0\in$~SR. Then one could apply the anomaly score to $x'$ and mitigate the background sculpting issue. We thank B.~Nachman for this suggestion. We also observe that a similar map $x\to x'$ is available directly from the \curtains{} method, without having to pass through the latent space; this could be used to prevent background sculpting in the \curtains{} method.} 

Furthermore, since $f$ is invertible, and likelihood ratios are invariant under coordinate reparametrizations, the $z$-space classifier should in principle be as asymptotically optimal as the $x$-space classifier, i.e.\
\beq
R(x)={p_{data}(x)\over p_{bg}(x)}={p_{data}(z)\over p_{bg}(z)}=R(z)
\eeq

The performance of the \cathode{} and \lacathode{} methods similarly rely on the quality of the trained and interpolated normalizing flow. For \cathode{} it controls the fidelity of the $p_{bg}(x)$ estimate, whereas for \lacathode{} the flow is responsible for $p_{data}(z)$. If the background events in data were not mapped to a unit normal distribution, both the learning of the likelihood ratio via the classifier and the decorrelation of auxiliary features from the resonant one would deteriorate.

We will show with examples in the following sections that \lacathode{} retains much of the excellent signal sensitivity as \cathode{}, while avoiding the sculpting of the $m$ distribution in the presence of correlations.\footnote{Another minor advantage of mapping the SR data to the latent space for the classification task is that the values of $m$ for this transformation are directly available, which simplifies things somewhat. In the case of \cathode{}, the mapping from the latent space samples to the data space via $f^{-1}(z;m)$ needs a separate estimation of the SR $m$ density.}

While \lacathode{} seems to be the superior enhanced bump hunting method, 
all is not lost for original \cathode{}---it remains a robust and powerful anomaly detection method as long as the correlations are sufficiently small. This is the case for the original feature set of the LHC Olympics 2020 R\&D dataset---as discussed in~\cite{Hallin:2021wme}, these have percent-level correlations with $m$, and we showed there that \cathode{} signal sensitivity remains robust to this small correlation (unlike \cwola{}-Hunting, which is more fragile to correlations). In this paper we demonstrate that the $m$ distribution is also not sculpted after a cut on $R(x)$.

\section{Toy model}
\label{sec:toy}

We begin by demonstrating the idea of \lacathode{} with a simple 1+2D toy model. In the first part, we will investigate how correlations between $x$ and $m$ affect the sculpting of $m$, and in the second part we will see what happens in the latent space. 

\begin{figure}[htb!]
    \centering
    \includegraphics[width=0.45\textwidth]{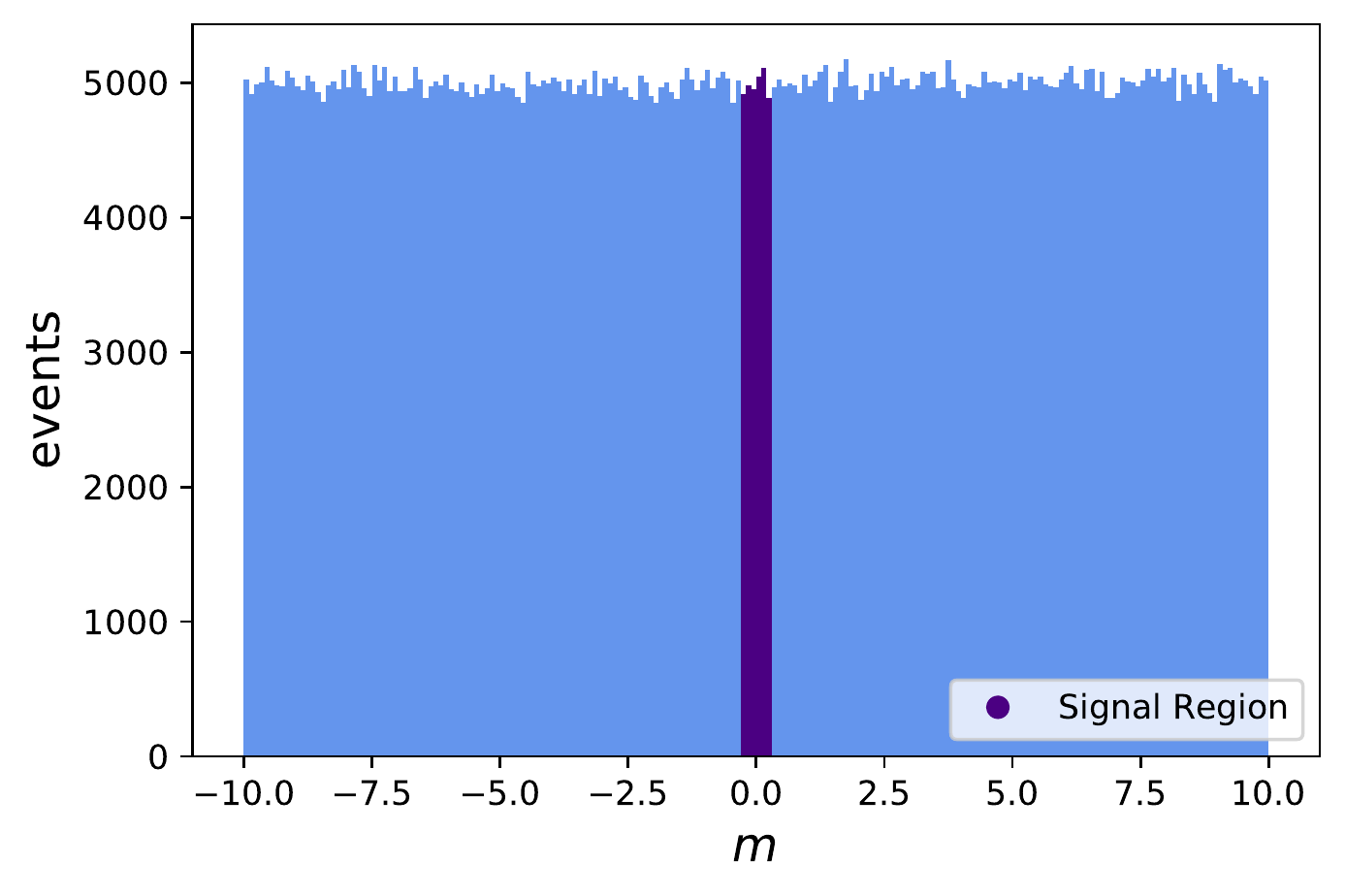}
    \caption{Uniform distribution of $m$, with signal region indicated in the darker band.}
    \label{fig:toy_model_m_distribution}
\end{figure}

The $m$ distribution will be sampled uniformly in the range $[-10,10]$. We will take the SR to be $m\in [-0.3,0.3]$. The $m$ distribution together with the SR is shown in Fig.~\ref{fig:toy_model_m_distribution}.

The features $x$ will be sampled from ${\mathcal N}(\mu=c\times m,\sigma=1)^2$. The parameter $c$ controls the amount of correlation between $x$ and $m$. 
We will consider $c~\in~\{0.001,0.1\}$.  

We generate two such sets, one ``data'' and one ``sample''. Of the 1 million events generated in each set, half are reserved for training, 1/6 for validation, and the remaining events are used to evaluate the trained classifier.

A binary classifier is trained in the SR to distinguish ``data'' from ``samples'' in $x$ space.\footnote{The classifier is implemented using Keras~\cite{chollet2015keras} with a TensorFlow~\cite{tensorflow2015-whitepaper} backend. It has three hidden layers with 64 nodes each and uses the optimizer Adam~\cite{kingma2014adam} with a learning rate of $10^{-3}$. Binary cross entropy is used for the loss function. It is trained for 50 epochs with a batch size of 128. The predictions of the five epochs with the lowest validation loss are ensembled to form an average prediction.} We find the cut values $R(x)>R_c$ that keep only the $1\%$  most anomalous events in the SR. 

Although only trained in the SR,
data on the entire interval $m\in [-10,10]$ are passed through the classifier and subject to the cut $R(x)>R_c$. If the classifier is not sculpting, it should return an $m$ distribution that looks uniform.

\begin{figure}[htb!]
    \centering
    \includegraphics[width=0.45\textwidth]{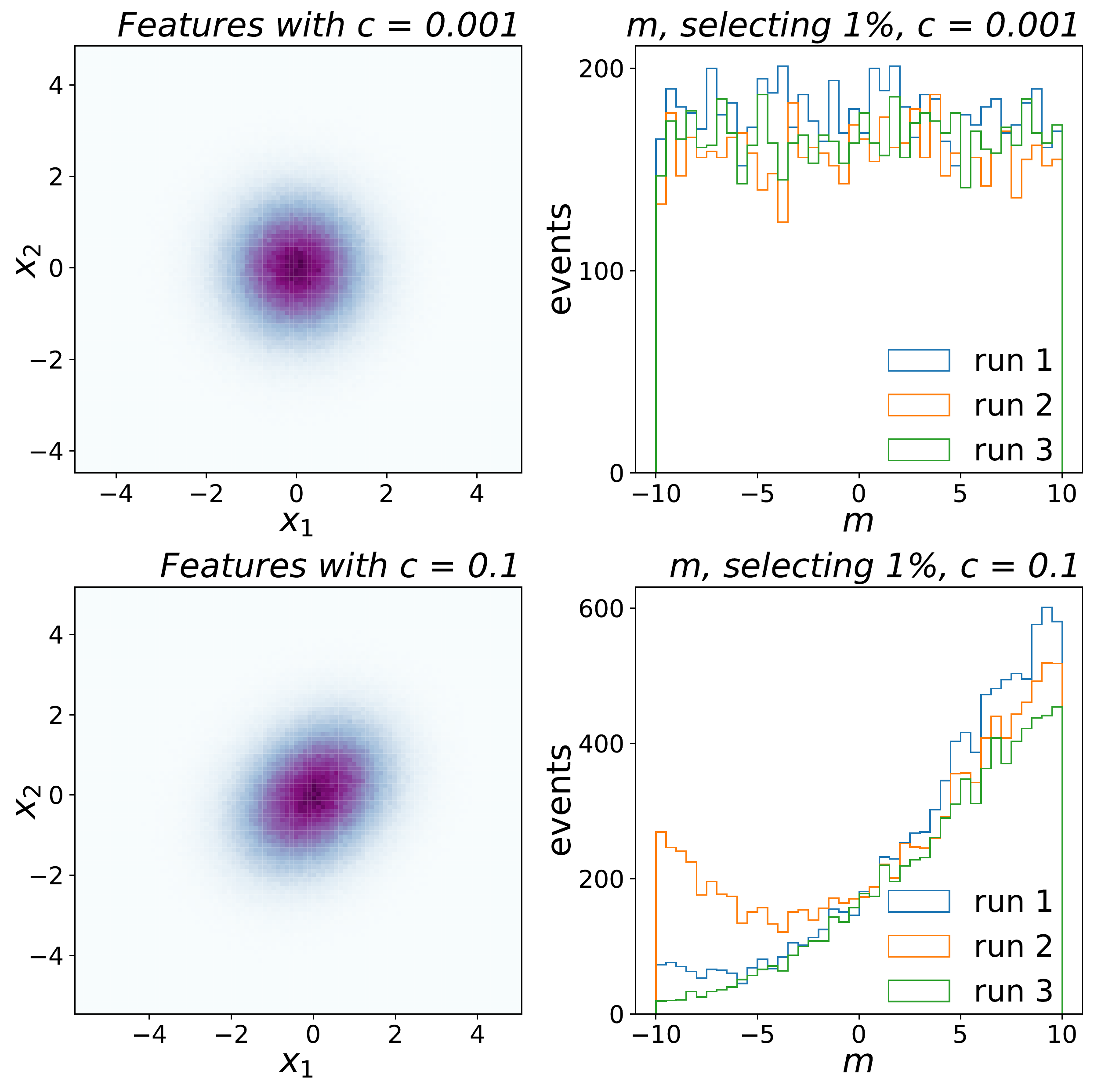}
    \caption{Features $x$ (left column) and the resulting $m$ distribution after a cut on the classifier score as described in the text (right column) under different correlations. The top row shows the case of a very small correlation, while the bottom row shows the features and results under a larger correlation.}
    \label{fig:toy_model_result}
\end{figure}

However, in the correlated case this is  not necessarily what happens. Shown in the right column of Fig.~\ref{fig:toy_model_result} are the $m$ distributions after cuts on the classifier, for different values of the correlation $c$ and for three independently trained classifiers on the same toy dataset. 
If the correlation is very small ($c=0.001$), no sculpting is seen. Meanwhile, if the correlation is sufficiently large ($c=0.1$), we see a severe sculpting in $m$. In this case, the $x$ distributions in the SB can be out of distribution relative to those in the SR, as seen in the left column of Fig.~\ref{fig:toy_model_result}. This can lead to unpredictable effects on the $m$ distribution after a cut on $R(x)>R_c$. 

Next we turn to the latent space, which will be an illustration of the \lacathode{} concept using this analytic toy model. Here we assume a perfect normalizing flow, so the transformation to the latent space is just
\beq
z=x-cm.
\eeq

\begin{figure}[htb!]
    \centering
    \includegraphics[width=0.45\textwidth]{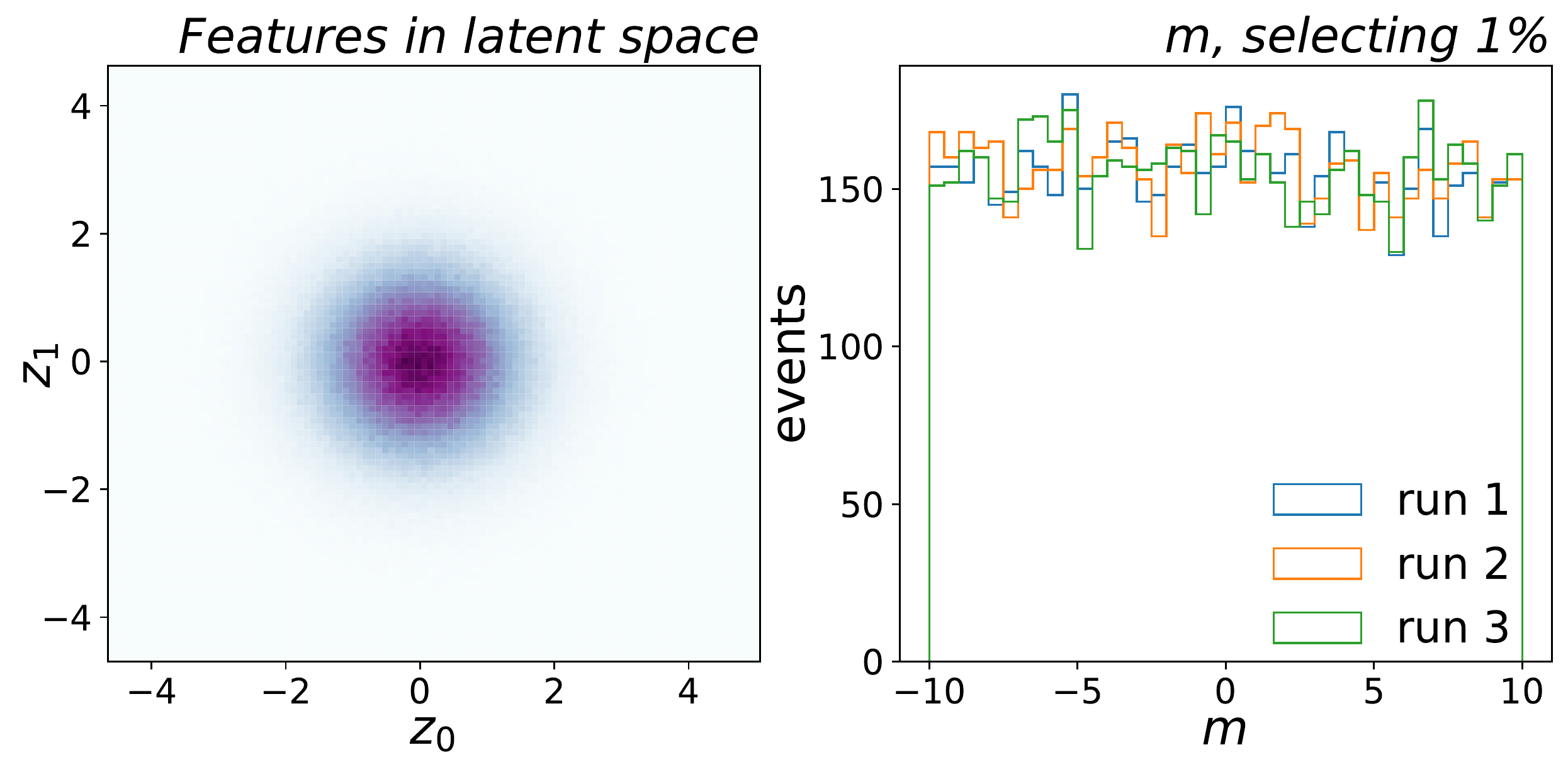}
    \caption{Features $z$ in latent space (left) and the resulting $m$ distribution after a cut on the classifier score as described in the text (right).}
    \label{fig:toy_model_result_noncorr}
\end{figure}

As can be seen in Fig. \ref{fig:toy_model_result_noncorr}, a classifier trained on the latent space data vs.\ samples will not show any sculpting in $m$.\footnote{We also point out that, through a quirk of this toy model, the latent space is equivalent to the $c=0$ case, so Fig.\ \ref{fig:toy_model_result_noncorr} also illustrates what happens in data space in the absence of any correlations.} We conclude that, as must be the case, \lacathode{} eliminates the sculpting of $m$ after a cut on $R(z)>R_c$. This of course makes perfect sense since $z$ and $m$ are completely independent of one another.

\section{LHCO R\&D dataset}

\label{sec:LHCO}

Let us now consider a demonstration of \lacathode{} with the LHC Olympics R\&D dataset~\cite{Kasieczka:2021xcg,gregor_kasieczka_2019_6466204}, where the resonant variable $m$ will be the dijet invariant mass $m_{JJ}$. 
For a description of the dataset and training details, refer to Appendix~\ref{sec:technical}.
Additional information about the dataset can be found in~\cite{Kasieczka:2021xcg,Hallin:2021wme}.

\subsection{Original features}

To set the stage, here we demonstrate \cathode{} and \lacathode{} with the original feature set $x=(m_1,\Delta m,\tau_{21}^{J_1},\tau_{21}^{J_2})$ of~\cite{Nachman:2020lpy,Hallin:2021wme}. 

\begin{figure*}[htb!]
    \centering
    \includegraphics[width=0.49\textwidth]{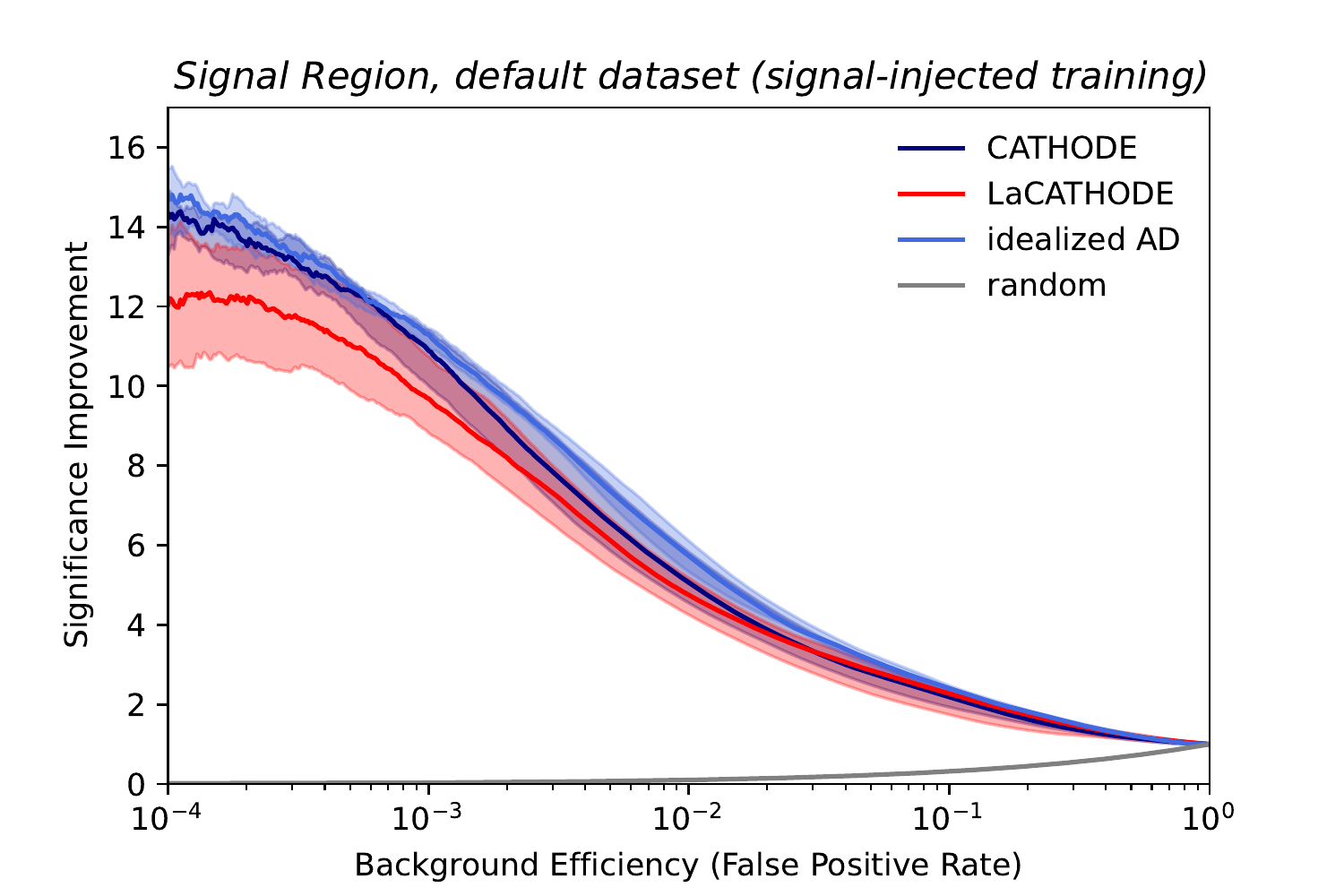}
    \includegraphics[width=0.49\textwidth]{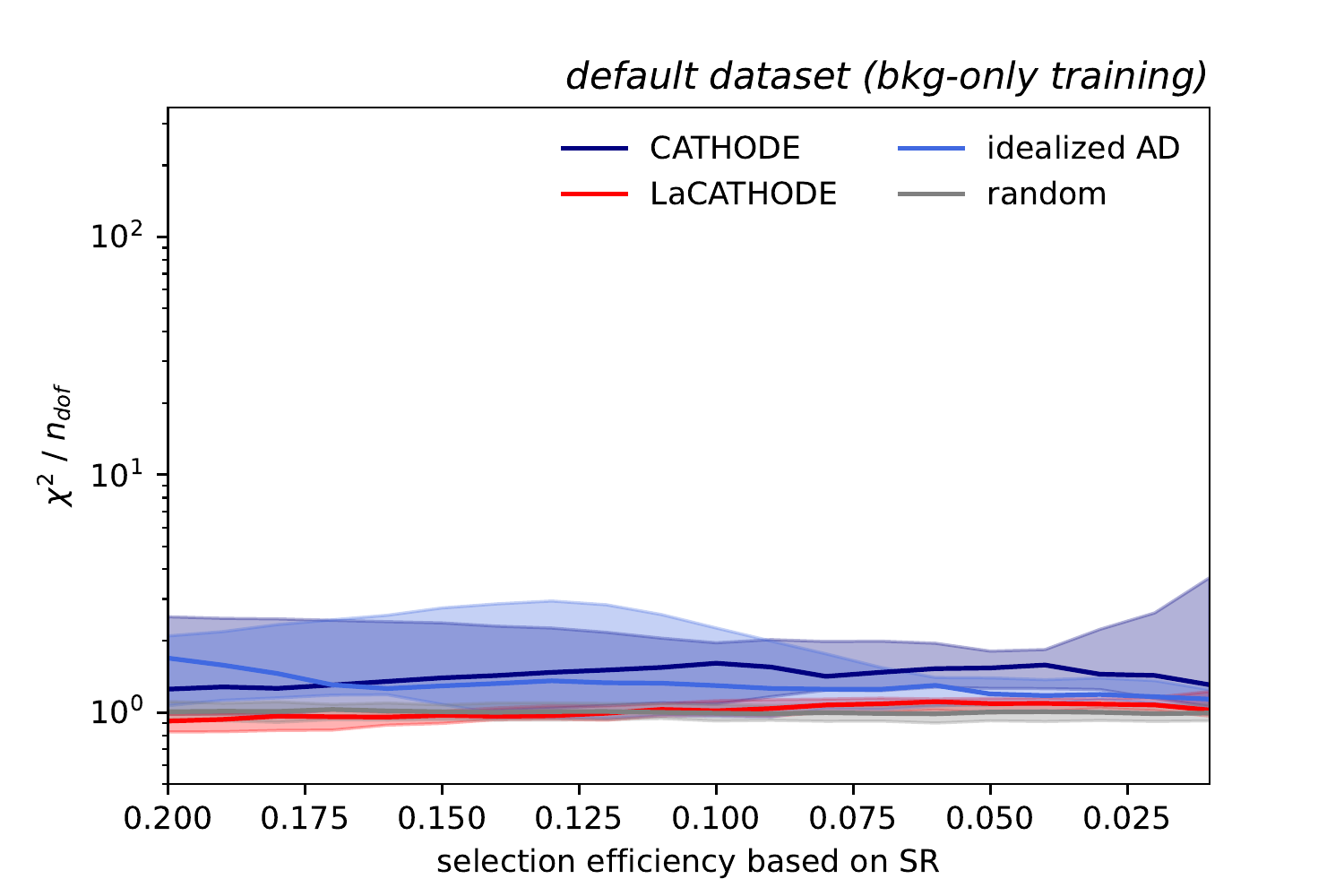}
    \includegraphics[width=0.49\textwidth]{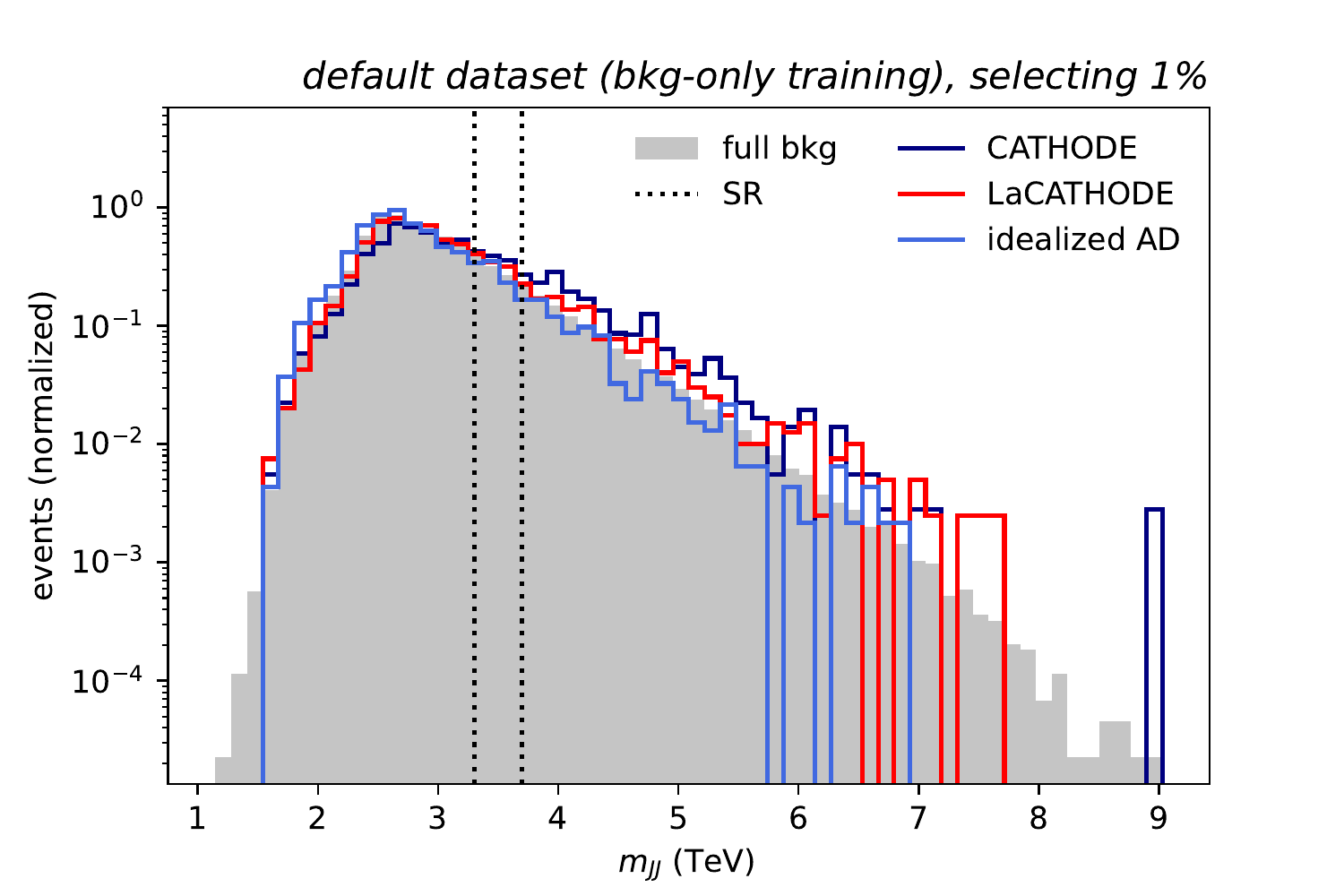}
    \caption{Performance comparison between \cathode{}, \lacathode{} and the idealized anomaly detector on the default dataset.
    The top left plot shows significance improvement in the SR as a function of background efficiency when training on the dataset with signal injected. The lines denote the median of ten independent retrainings whereas the bands show the central 68\%.
    Similarly, the top right plot shows the distance between the distribution before and after a selection, quantified by the median and central 68\% of the $\chi^2 / n_{dof}$ metric, at continuous choices of selection efficiency. The denoted efficiency is based on the SR and the same anomaly score threshold is applied to the SB. In this case, the anomaly detectors were trained on pure background.
    The bottom plot shows the resulting normalized $m_{JJ}$ distributions, after selecting the 1\% most anomalous events, based on the first of the ten background-only trainings, overlayed with the distribution before a cut.
    Overall, we see that for the original feature set, where correlations between the features and $m_{JJ}$ are small, the sculpting in original \cathode{} and the idealized AD in $x$ space are reasonably under control, although the variance in the $\chi^2$ distance metric is higher than for \lacathode{}.}
    \label{fig:lhco_performance_default}
\end{figure*}

First, we turn to Fig.~\ref{fig:lhco_performance_default} (top left), which shows the significance improvement characteristic (SIC) curves using the same injection as in~\cite{Hallin:2021wme}, in terms of the median and the central 68\% bands of ten independent trainings. Recall that the SIC is defined as the improvement in the nominal significance, $\epsilon_S/\sqrt{\epsilon_B}$, where $\epsilon_S$ and $\epsilon_B$ are the  signal and background efficiencies in the SR, respectively, after a cut on the anomaly score. Unlike in~\cite{Hallin:2021wme}, here we show the SIC as a function of the background efficiency. We see from Fig.~\ref{fig:lhco_performance_default} (top left) that \lacathode{} retains much of the signal sensitivity as original \cathode{}.

Shown in Fig.~\ref{fig:lhco_performance_default} (bottom) is the $m_{JJ}$ distribution for the original feature set for background events after cuts on $R(x)$, corresponding to keeping only the 1\% most anomalous events within the SR.\footnote{Based on the SIC curves, one would be inspired to use a working point with tighter background efficiency, e.g. 0.1\% or even 0.01\%. However, the 1\% selection efficiency choice here is made because of the limited size of the test set.} By eye, we see that there is only minimal sculpting. 

In order to quantify this, we compute a histogram from the test data $m_{JJ}$ before selection corresponding to $n_{bins} = 300$ bins with equal bin content ($N_{i}^{full} = N^{tot}/n_{bins}$ for each bin $i$) and normalize it to unit area $n_{i}^{full} = a_{i} N_{i}^{full}$. Here $n_{bins}$ has been chosen such that a 1\% selection still retains at least ten test set events in every bin. After applying a selection of $\varepsilon_{sel}$ using an anomaly detector, the resulting $m_{JJ}$ is histogrammed with the same binning $N_{i}^{sel}$ and also normalized to unit area $n_{i}^{sel} = b_{i} N_{i}^{sel}$. One then obtains a distance metric $\chi^2$ as a function of decreasing $\varepsilon_{sel}$:
\begin{equation} \label{eq:chi2}
    \chi^2 = \sum_{i} \frac{\left( n_{i}^{full} - n_{i}^{sel} \right)^2}{ \sigma_{i}^2 } = \sum_{i} \frac{\left( n_{i}^{full} - n_{i}^{sel} \right)^2}{ a_{i}^2 \varepsilon_{sel} N_{i}^{full} }.
\end{equation}

For better reference, the resulting $\chi^2$ is divided by the number of degrees of freedom $n_{dof} = n_{bins} - 1$. This quantity is shown for \cathode{}, \lacathode{} and the idealized anomaly detector in Fig.~\ref{fig:lhco_performance_default} (top right), again in terms of the median and the central 68\% bands of ten independent trainings. While all methods have values close to unity, both \cathode{} and the idealized anomaly detector have a higher variance than \lacathode{}, which overlaps for the most part with the $\chi^2$ that one obtains from 100 randomly drawn subsets of events with the target selection efficiency.

As in the toy example, this shows that \cathode{} can be robust against sculpting for small enough correlations. This means that as long as the features are well chosen, original \cathode{} can be a viable enhanced bump hunt method on its own, and superior to \cwola{}-Hunting, which is more sensitive to correlations.

\subsection{Shifted features}

\begin{figure*}[htb!]
    \centering
    \includegraphics[width=0.49\textwidth]{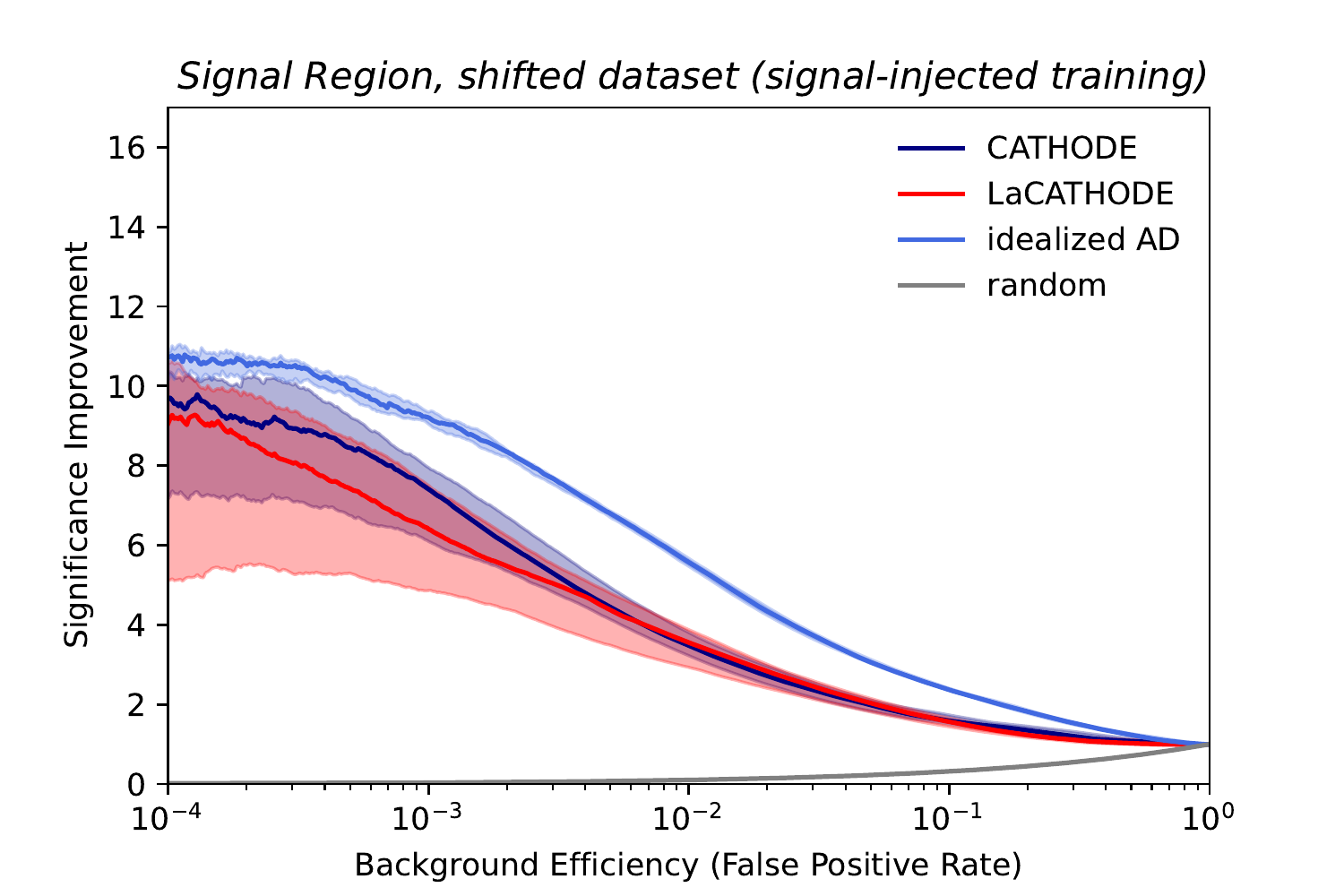}    \includegraphics[width=0.49\textwidth]{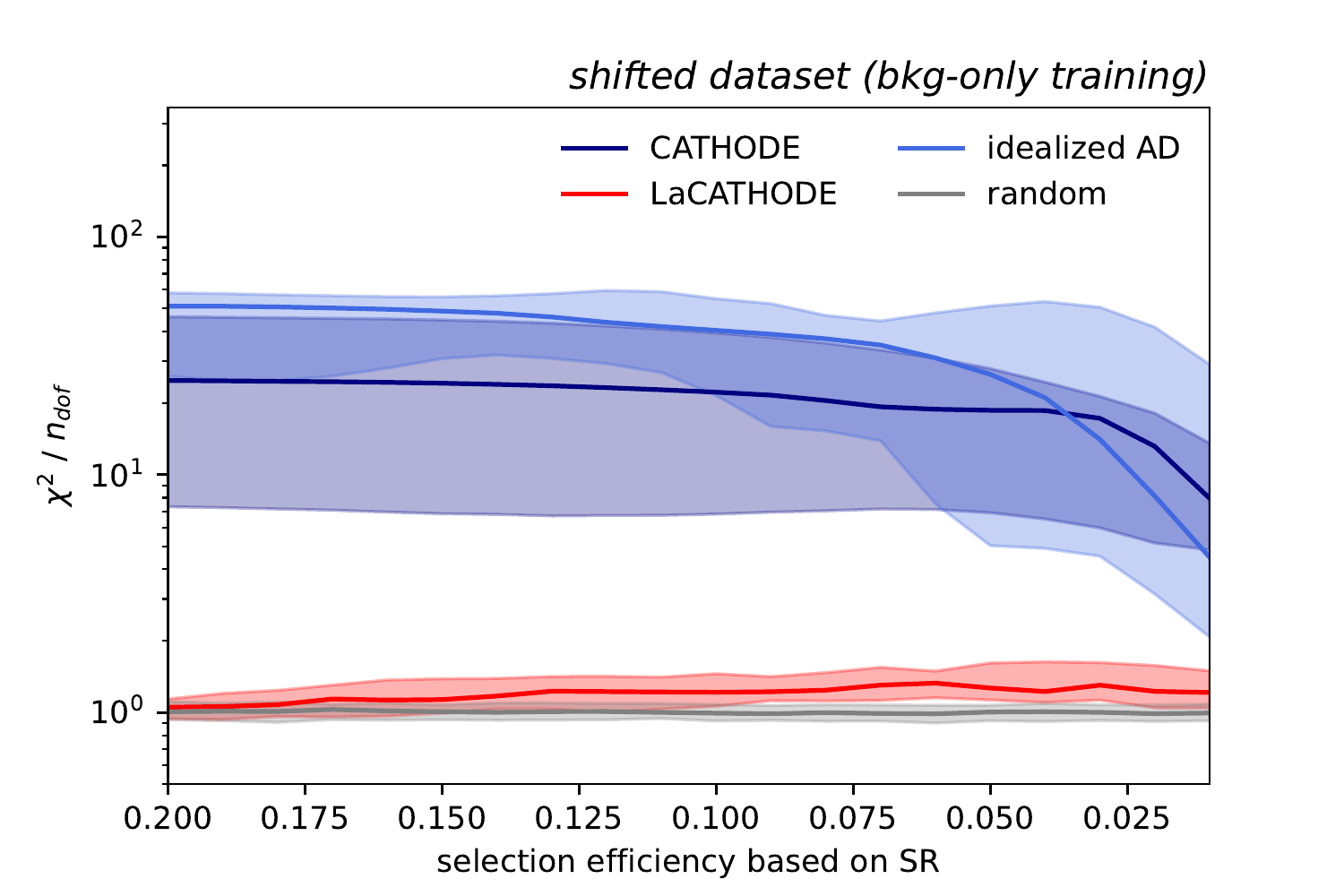}
    \includegraphics[width=0.49\textwidth]{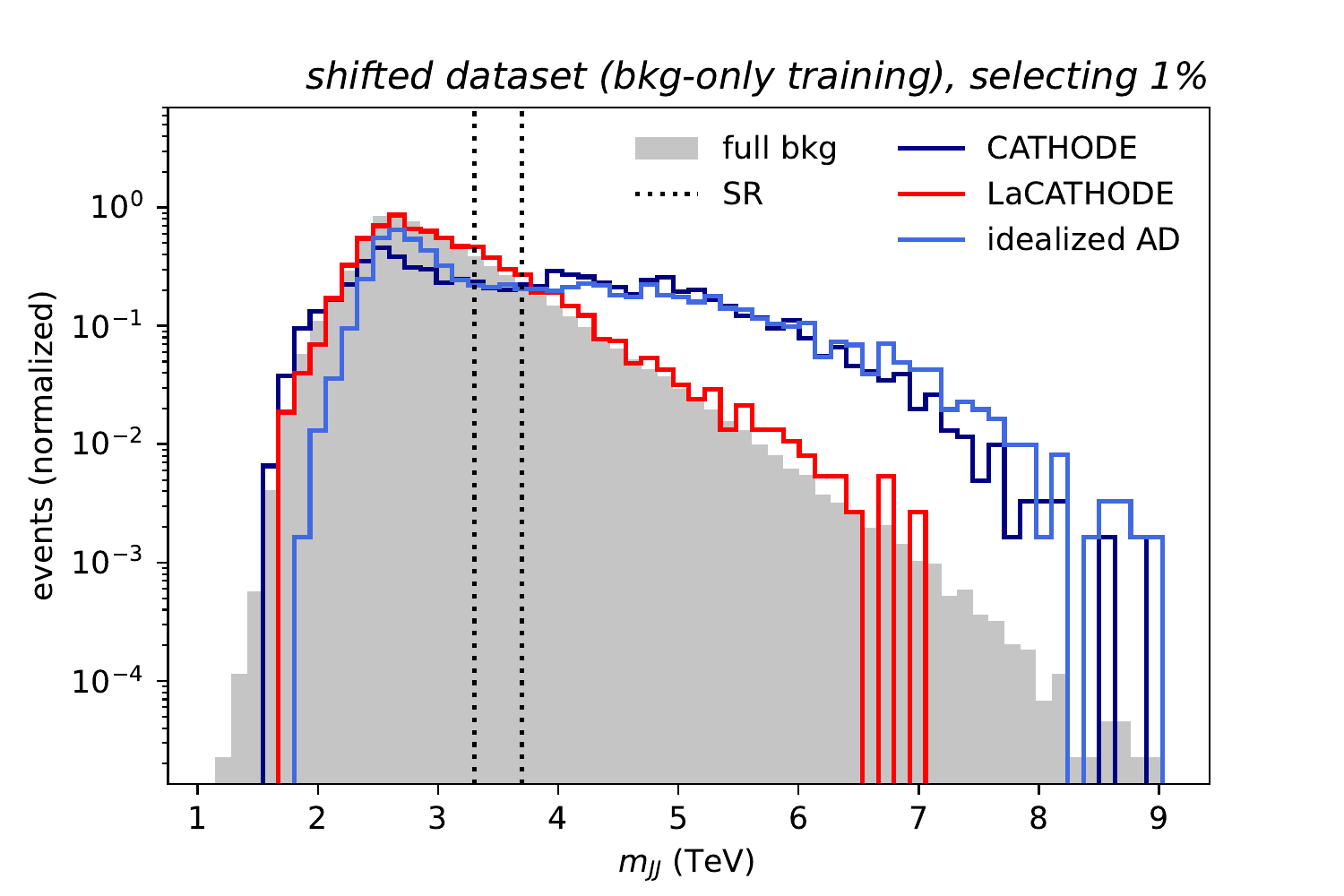}
    \caption{The same performance comparison between anomaly detectors as in Fig.~\ref{fig:lhco_performance_default}, but trained and evaluated on the shifted dataset $x'=(m_1+cm_{JJ},\Delta m+cm_{JJ},\tau_{21}^{J_1},\tau_{21}^{J_2})$ with $c=0.1$. Here we see much more significant sculpting in original \cathode{} and the idealized AD trained on $x$ space, while \lacathode{} is again completely stable to the cut on the anomaly score.}
    \label{fig:lhco_performance_shifted}
\end{figure*}

Next, we show the sculpting with the shifted feature set $x'=(m_1+cm_{JJ},\Delta m+cm_{JJ},\tau_{21}^{J_1},\tau_{21}^{J_2})$ where we choose $c=0.1$ as in~\cite{Nachman:2020lpy,Hallin:2021wme}. Now we observe in Fig.~\ref{fig:lhco_performance_shifted} that after cuts on $R(x)$ the sculpting is quite severe, again as expected from the toy model.

For \lacathode{} on the other hand, sculpting is still almost nonexistent after cuts on $R(z)$. This confirms what we found with the toy model (using now a trained normalizing flow and not a perfect one): the normalizing flow essentially decorrelates $z$ and $m_{JJ}$ and solves the problem of sculpting in the $m_{JJ}$ distribution.

Figure~\ref{fig:lhco_performance_shifted} (top left) shows the SIC curves for \lacathode{} for the LHCO R\&D signal with shifted features; we see that \lacathode{} more or less matches the original \cathode{}.

\subsection{Including $\Delta R$}

Finally, we demonstrate the performance of \cathode{} and \lacathode{} in a somewhat more well-motivated feature set with correlations, i.e.\ including $\Delta R$. The angular distance $\Delta R$ between the two jets is defined as:
\beq
\Delta R = \sqrt{(\phi_{J_{2}} - \phi_{J_{1}})^{2} + (\eta_{J_{2}} - \eta_{J_{1}})^{2}}
\eeq

This feature was first suggested in~\cite{Raine:2022hht} as it might be relevant for certain signal models (e.g.\ due to different initial state radiation patterns compared to QCD jets) and because its strong correlation with the invariant mass introduces a well-motivated test of anomaly detection in the presence of correlations.

\begin{figure*}[htb!]
    \centering
    \includegraphics[width=0.49\textwidth]{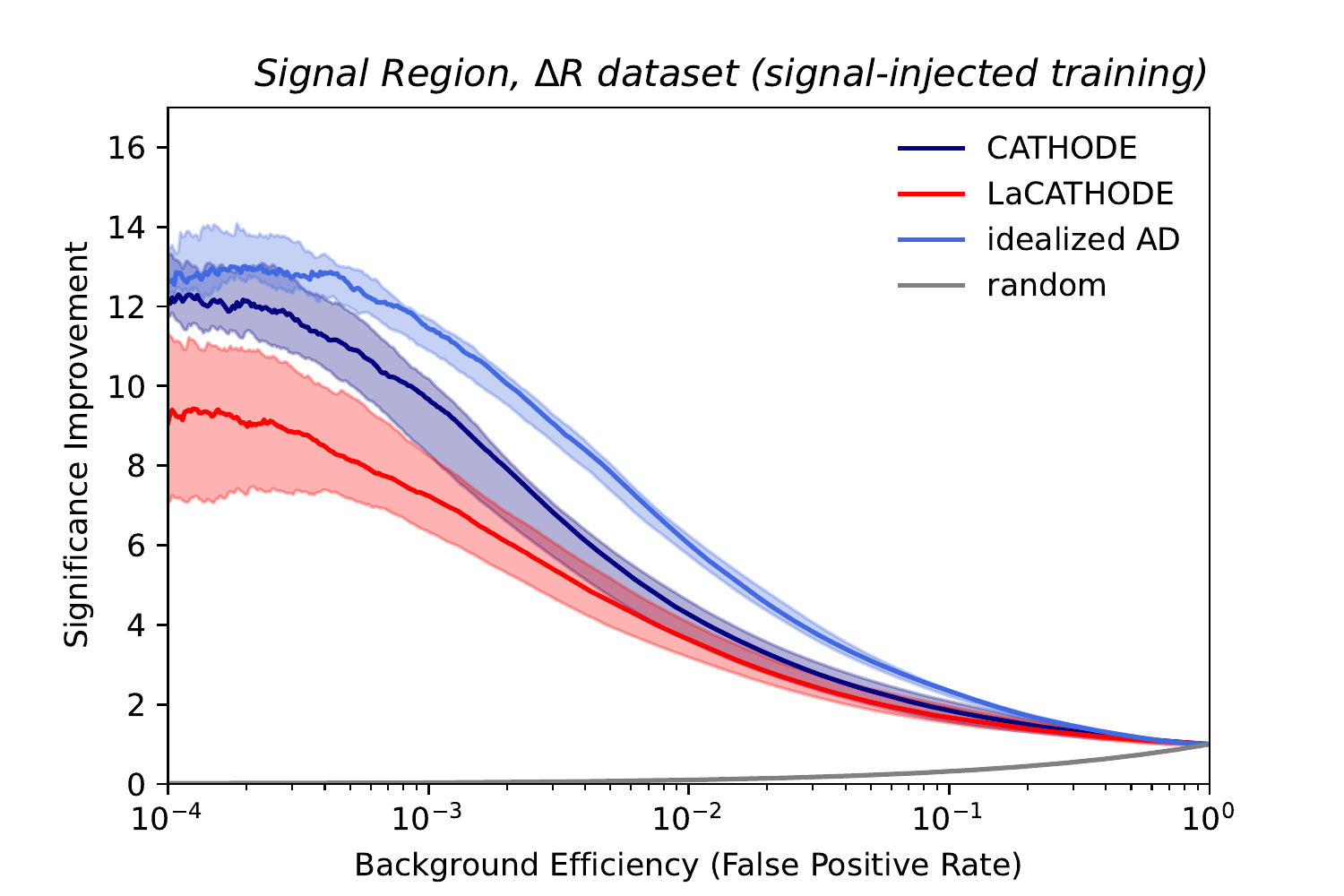}
    \includegraphics[width=0.49\textwidth]{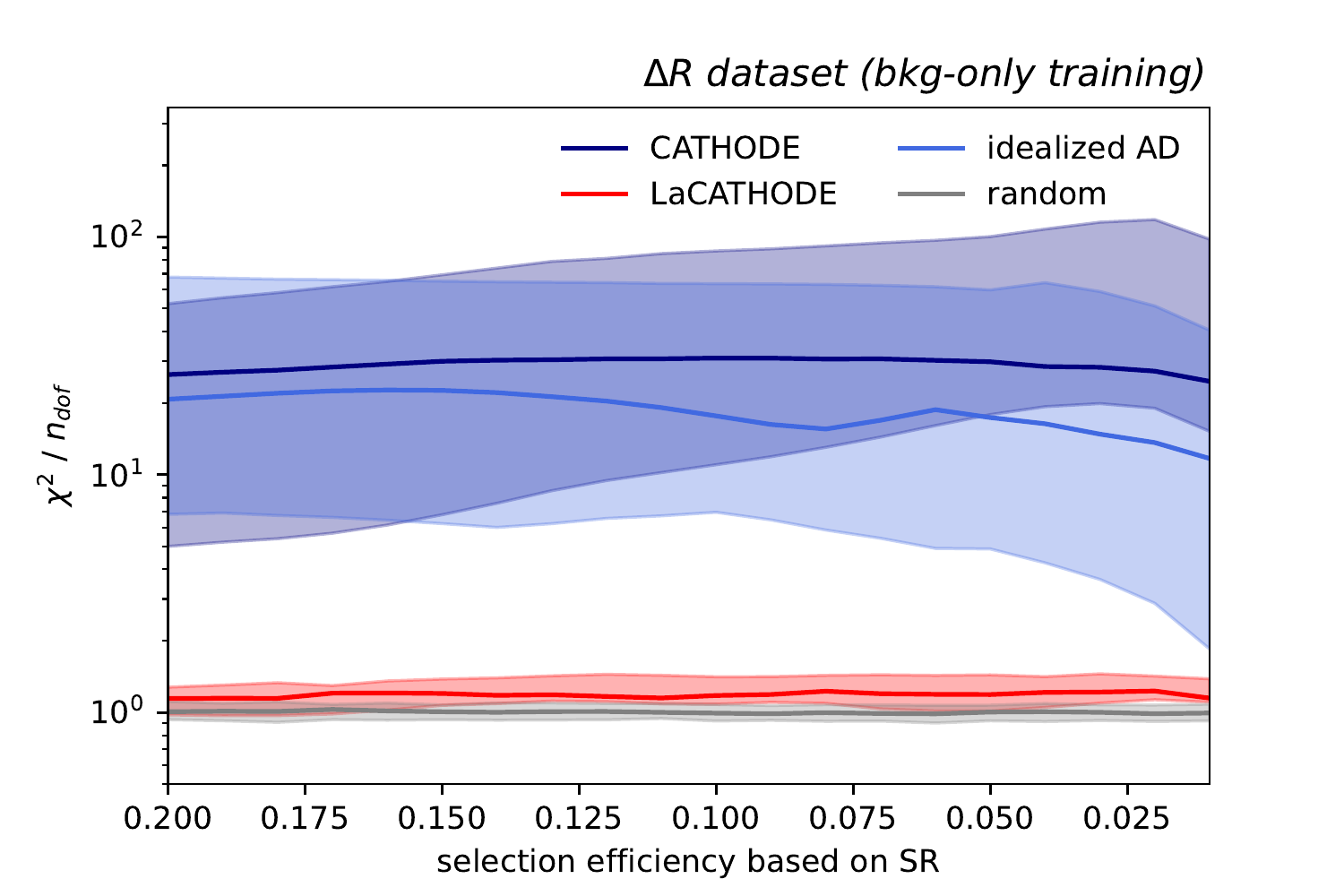}
    \includegraphics[width=0.49\textwidth]{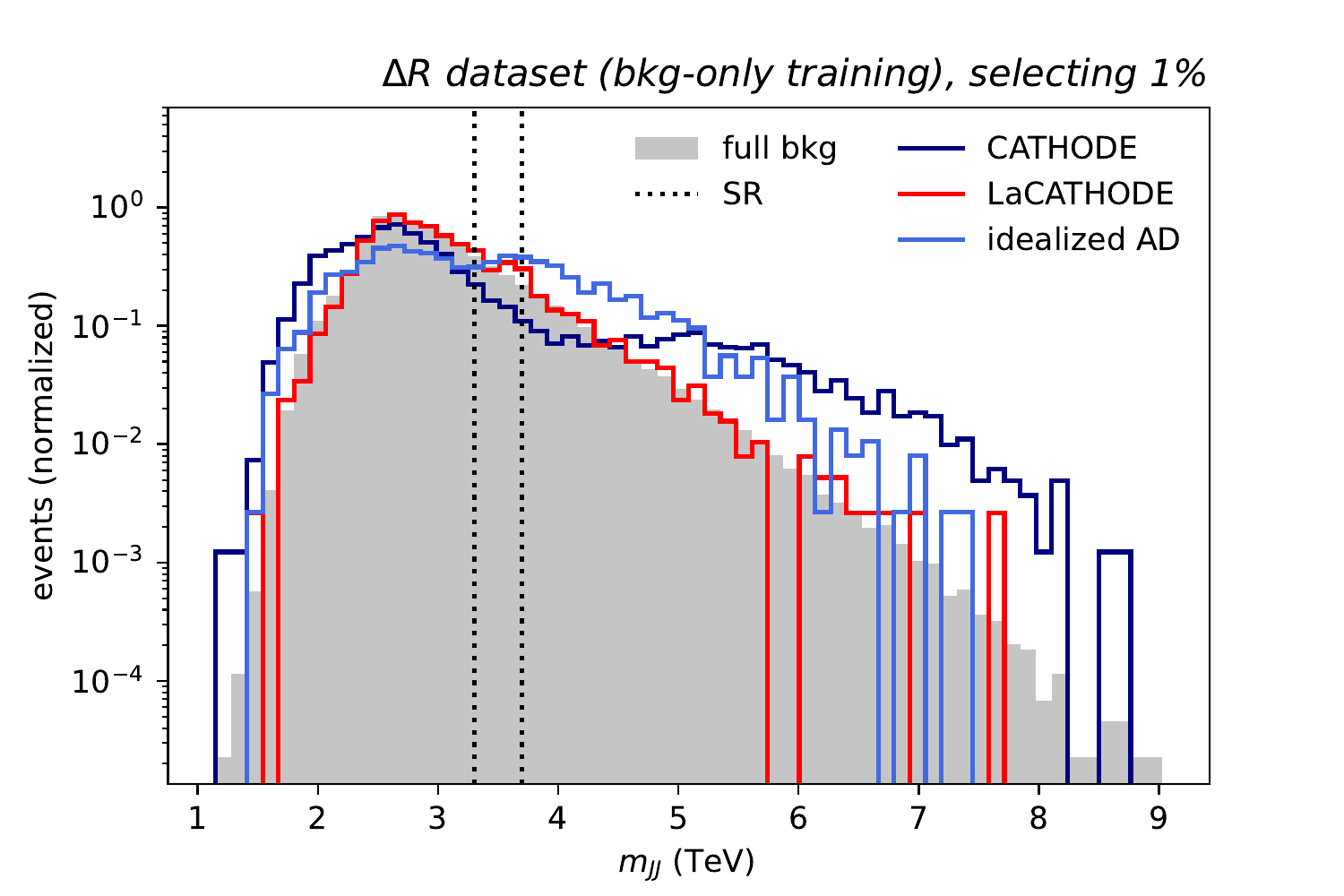}
    \caption{The same performance comparison between anomaly detectors as in Fig.~\ref{fig:lhco_performance_default}, but trained and evaluated on the dataset $x''=(m_1,\Delta m,\tau_{21}^{J_1},\tau_{21}^{J_2}$, $\Delta R)$. As in the case with the artificial correlations, we again see much more significant sculpting in original \cathode{} and the idealized AD trained on $x$ space, while \lacathode{} is once again completely stable to the cut on the anomaly score.}
    \label{fig:lhco_performance_deltar}
\end{figure*}

We see in Fig.~\ref{fig:lhco_performance_deltar} that \cathode{} has severe sculpting after cutting on $R(x)$, while \lacathode{} has essentially no sculpting. We see that in this case the signal sensitivity of \lacathode{} is lower than that of \cathode{}, however, in a real-world setting this might be an acceptable trade-off for having a proper background estimation.

\section{Conclusions}
\label{sec:conclusions}

In this work, we have identified and addressed a problem with ``enhanced bump hunt"--type anomaly detectors: the unwelcome sculpting of background distributions when selecting events based on an anomaly  score, in the presence of correlations between input features and the primary resonant feature. 

A large amount of background sculpting in the distribution of the primary resonant feature can make the downstream analysis much more difficult or impossible. For example, 
the resulting structures (e.g. peaks or enhanced tails) could be mistaken for spurious signals. Alternatively, the process of fitting the complicated, sculpted distribution introduces additional model assumptions (e.g.\ simulation dependence), or makes the fit function overly generic and increases the uncertainty on the background prediction. 
An unsculpted distribution instead allows one to obtain the background template from the inclusive spectrum and preserves the fully data-driven nature of the enhanced bump hunt.

After showing in an analytic toy example how increased correlations between features can result in such sculpting, we analyzed the LHC Olympics R\&D dataset, using the recently proposed \cathode{} method for enhanced bump hunting. In this more physically motivated context, we confirmed that \cathode{} indeed results in strong sculpting of the $m$ distribution when using highly correlated features.

Our proposed solution to the problem, named La(tent)\cathode{}, maintains a similar performance as measured by the significance improvement characteristic, but strongly suppresses the unwanted shaping of the background $m$ distribution.
This is achieved by using the conditional normalizing flow trained on sideband data that underlies the \cathode{} method. This SB density estimator maps background data everywhere (not just in the signal region) to a latent space that is normally distributed, for every $m$. Thus the anomaly score learned in the latent space becomes essentially decorrelated from $m$ in the background, and the problem of sculpting is eliminated.

While yielding an important gain in stability, the modified \lacathode{} method is conceptually and computationally no more complex than the original \cathode{} method and we look forward to its swift experimental deployment.

\section*{acknowledgements}

We thank the organizers of the ``A Deep-Learning Era of Particle Theory'' workshop at MITP Mainz and the organizers of the ``Hammers \& Nails 2022'' workshop at Weizmann Institute of Science for their kind hospitality and for providing the opportunity for exchange that formed the basis of this work. We are grateful to Ben Nachman for feedback on the draft.
GK, TQ, and MS acknowledge support by the Deutsche Forschungsgemeinschaft under Germany’s Excellence Strategy – EXC 2121  Quantum Universe – 390833306.\ 
The work of TQ and MS was supported by BMBF grant 05H21GUCC1.
The work of AH and DS was supported by DOE grant DOE-SC0010008.\  This research was supported in part through the Maxwell computational resources operated at Deutsches Elektronen-Synchrotron DESY, Hamburg, Germany.

\FloatBarrier
\appendix

\section{Technical details}

\label{sec:technical}

\subsection{Dataset specifications}

The original LHCO R\&D dataset~\cite{gregor_kasieczka_2019_6466204} consists of a total of 1 million QCD dijet events that constitute the Standard Model (SM) background and 100,000 events of a beyond the Standard Model (BSM) signal process. The signal model used is $Z'\rightarrow XY$ with hadronically decaying daughter particles $X\rightarrow q\bar{q}$ and $Y\rightarrow q\bar{q}$ and respective masses of $m_{Z'}=3.5~$TeV, $m_{X}=500~$GeV and $m_{Y}=100~$GeV. 

The events were produced with  \texttt{Pythia 8.219}~\cite{Sjostrand:2006za,Sjostrand:2007gs} and \texttt{Delphes 3.3.1}~\cite{deFavereau:2013fsa,Mertens:2015kba,Selvaggi:2014mya} using default settings and without inclusion of pileup or multiparton interactions. Event selection is applied using a trigger requiring a single large-radius jet ($R=1$) clustered with the anti-$k_{\mathrm{T}}$ algorithm~\cite{Cacciari:2008gp} and passing a $p_{\mathrm{T}}$ threshold of 1.2 TeV. Jet clustering was performed by \texttt{Fastjet}~\cite{Cacciari:2011ma,Cacciari:2005hq}.

To form the ``data" (a proxy for real, unlabeled LHC data) used in this study, we 
injected 1000 signal events into the 1M QCD background events. This corresponds to a signal-to-background ratio of $0.6\%$ and an initial nominal significance of $\frac{S}{\sqrt{B}}=2.2$ inside the signal region.

One difference compared to the original \cathode{} studies is here we have simplified the split of the ``data" into training (1/2), validation (1/6) and test sets (1/3). The test set is set aside and used only for final evaluation plots (i.e.\ plots of the $m_{JJ}$ distribution, SIC curves, etc.). 

For the classifier part of \cathode{}/\lacathode{}, 267,000 synthetic background events are sampled, either from the conditional normalizing flow that has been interpolated into the SR (\cathode{}), or from the standard normal distribution (\lacathode{}). Following the train/val split above, 3/4 of these events (200,000) are used for training the classifier and 1/4 (67,000) are used for validation. 

As in the original \cathode{} work~\cite{Hallin:2021wme} we also use an additional QCD background sample for studies of an idealized anomaly detector \cite{david_shih_2021_5759087}. It consists of 612,858 background events in the signal region and is produced using the exact same procedure as the QCD events from the LHCO R\&D dataset. To have a one-to-one comparison with \cathode{} and \lacathode{} results, we also use 200,000 SR background events of the additional production for training and 67,000 SR background events for validation.

Finally, for the evaluation we distinguish two studies: significance improvement and sculpting. For the significance improvement studies, we use all of the available remaining signal and background events that were not used in either the training or the validation sets before. These same events are used for all methods such that the results are comparable. Meanwhile, for the sculpting studies, only background events are considered, this time, however, both in the SR and in the SB. Here, we use the test set of the original LHCO R\&D dataset, which contains 333,000 background events. The respective numbers of signal and background events for the different methods and datasets are summarized in Table~\ref{tab:datacount}.

\subsection{Training details}

In terms of the technical implementation of the neural networks, we follow strictly the setup of Ref.~\cite{Hallin:2021wme}. This holds for the architecture of normalizing flow and classifier as well as their training and how the predictions of the 10 classifier training epochs with lowest validation loss are ensembled. One difference, however, is the omission of this type of ensembling in the case of the normalizing flow. While this 10-epoch ensembling was previously realized for \cathode{} as drawing one tenth of the background-like sample from each of the 10 epochs with lowest validation loss, we leave it up to future studies how one would optimally perform such an ensembling when mapping from data space to latent space. For better comparison, we also omitted the sampling ensemble in the implementation of the \cathode{} benchmark in this paper.

\begin{table*}[!ht]
\caption{Number of events in training, validation and test datasets used for the investigated anomaly detection methods. Numbers are rounded to the nearest multiple of 1,000.} 
\label{tab:datacount}
\begin{center}
  \begin{tabular}{|c|c|c|c|c|}
    \hline
    Method & Type & Training & Validation & Evaluation\\
    \hline 
    \multirow{3}{*}{\cathode{}} & density estimator & 439k SB data & 147k SB data & \multirow{8}{*}{\makecell{For sculpting studies:\\333k background (SB+SR)\\\\For significance improvement:\\386k SR background\\75k SR signal}}\\
    \hhline{|~|-|-|-|~|}
    & \multirow{2}{*}{classifier} & 61k SR data & 20k SR data &\\
    & & 200k SR background samples & 67k SR background samples &\\
    \hhline{|-|-|-|-|~|}
    \multirow{3}{*}{\lacathode{}} & density estimator & 439k SB data & 147k SB data &\\
    \hhline{|~|-|-|-|~|}
    & \multirow{2}{*}{classifier} & 61k SR data in latent space & 20k SR data in latent space &\\
    & & 200k latent space samples & 67k latent space samples &\\
\hhline{|-|-|-|-|~|}
    \multirow{2}{*}{Idealized AD}  & \multirow{2}{*}{classifier}    & 61k SR data &  20k SR data &\\
    & & 200k SR background & 67k SR background &\\
    \hline
  \end{tabular}
\end{center}
\end{table*}

\subsection{Code}

The code to reproduce the results in this paper is provided at: \url{https://github.com/HEPML-AnomalyDetection/CATHODE/tree/LaCATHODE}.

\FloatBarrier
\bibliography{HEPML,other}
\bibliographystyle{apsrev4-1}

\end{document}